\documentclass[aps,twocolumn,showpacs,preprintnumbers,prl,superscriptaddress,groupedaddress,10pt]{revtex4-1}
\usepackage[utf8]{inputenc}
\usepackage{graphicx,amssymb,amsmath,amsthm,amsfonts,epsfig,times,natbib}

\usepackage[usenames,dvipsnames]{color}
\usepackage{epstopdf}
\usepackage{amsmath,amssymb}
\usepackage{tensor}
\usepackage{mathtools}
\usepackage{amsbsy}
\usepackage{bm,url}
\usepackage[linktocpage]{hyperref}

\usepackage[scaled]{beramono}
\usepackage[T1]{fontenc}

\definecolor{oxfordblue}{rgb}{0.0, 0.13, 0.28}
\definecolor{burgundy}{rgb}{0.5, 0.0, 0.13}
\definecolor{darkolivegreen}{rgb}{0.33, 0.42, 0.18}
\definecolor{darkblue}{rgb}{0,0,0.5}
\definecolor{richcarmine}{rgb}{0.84, 0.0, 0.25}
\definecolor{darkblue}{rgb}{0,0,0.5}
\definecolor{venetianred}{rgb}{0.78, 0.03, 0.08}
\definecolor{skobeloff}{rgb}{0.0, 0.48, 0.45}
\hypersetup{colorlinks=true, citecolor=darkblue, linkcolor=darkblue,
urlcolor = darkblue}

\def\apjl{\rm{ApJ}}
\def\apjs{\rm{ApJS}}
\def\ao{\rm{Appl.~Opt.}}

\def\aap{\rm{A\&A}}

\newcommand{\ben}{\begin{enumerate}}
\newcommand{\een}{\end{enumerate}}

\def\be{\begin{equation}}
\def\ee{\end{equation}}
\def\bea{\begin{eqnarray}}
\def\eea{\end{eqnarray}}

\newcommand{\beq}{\begin{eqnarray}}
\newcommand{\eeq}{\end{eqnarray}} 
\newcommand{\ba}{\begin{align}}
\newcommand{\ea}{\end{align}}

\begin{document}

\title{Light ring stability in ultra-compact objects}

\author{
Pedro V. P. Cunha$^{1,2}$,
Emanuele Berti$^{3,2}$,
Carlos A. R. Herdeiro$^{1}$
}

\affiliation{${^1}$ Departamento de F\'isica da Universidade de Aveiro and CIDMA,
Campus de Santiago, 3810-183 Aveiro, Portugal}
\affiliation{${^2}$ CENTRA, Departamento de F\'isica, Instituto Superior
T\'ecnico, Universidade de Lisboa, Avenida Rovisco Pais 1,
1049 Lisboa, Portugal}
\affiliation{${^3}$ Department of Physics and Astronomy, The University of 
Mississippi, University, MS 38677, USA}

\begin{abstract}
  We prove the following theorem: axisymmetric, stationary solutions
  of the Einstein field equations formed from classical gravitational
  collapse of matter obeying the null energy condition, that are
  everywhere smooth and ultracompact (i.e., they have a light ring)
  must have at least \textit{two} light rings, and one of them is
  \textit{stable}. It has been argued that stable light rings
  generally lead to nonlinear spacetime instabilities. Our result
  implies that smooth, physically and dynamically reasonable
  ultracompact objects are not viable as observational alternatives to
  black holes whenever these instabilities occur on astrophysically
  short time scales. The proof of the theorem has two parts: (i) We
  show that light rings always come in pairs, one being a saddle point
  and the other a local extremum of an effective potential. This
  result follows from a topological argument based on the Brouwer
  degree of a continuous map, with no assumptions on the spacetime
  dynamics, and hence it is applicable to any metric gravity theory
  where photons follow null geodesics. (ii) Assuming Einstein's
  equations, we show that the extremum is a local minimum of the
  potential (i.e., a stable light ring) if the energy-momentum tensor
  satisfies the null energy condition.
\end{abstract}

\maketitle

%%%%%%%%%%%%%%%%%%%%%%%%%%%%%%%%%%%%%%%%%%%%%%%%%%%%%%%%%%%%%%%%%%%%%%%%%%%%%
\noindent{{\bf{\em Introduction.}}}
%%%%%%%%%%%%%%%%%%%%%%%%%%%%%%%%%%%%%%%%%%%%%%%%%%%%%%%%%%%%%%%%%%%%%%%%%%%%%
%
The historic LIGO gravitational wave~(GW) detections~\cite{Abbott:2016blz,Abbott:2016nmj,Abbott:2017vtc} provide strong evidence that astrophysical black holes (BHs) exist and merge. LIGO and the space-based detector LISA~\cite{Audley:2017drz} will allow us to test the nature of compact objects and the strong-field dynamics of general relativity in unprecedented ways~\cite{Gair:2012nm,Yunes:2013dva,Berti:2015itd,TheLIGOScientific:2016src,Berti:2016lat}.

All LIGO detections so far are consistent with the inspiral, merger and ringdown waveforms produced by binary BH mergers. In particular, the ringdown phase is sourced by the relaxation of the final perturbed BH into equilibrium, and it has been regarded as a distinctive signature of BHs~\cite{Berti:2005ys,Berti:2009kk}. There is a well-known correspondence between the complex quasinormal oscillation frequencies of a BH and perturbations of the light ring~\cite{1972ApJ...172L..95G,Cardoso:2008bp,Berti:2014bla,Glampedakis:2017dvb}. Intriguingly, because of this correspondence, all compact objects with a {circular photon orbit, i.e. a light ring (LR),} but with no horizon -- hereafter dubbed \textit{Ultra-Compact Objects} (UCOs) -- initially vibrate like BHs, and only later display oscillations features that depend on their internal structure ($w$-modes or ``echoes''~\cite{1991RSPSA.434..449C,Kokkotas:2003mh,Ferrari:2000sr,Damour:2007ap,Barausse:2014tra,Cardoso:2016rao,Cardoso:2016oxy,Mark:2017dnq}). Therefore LIGO observations of a ringdown signal consistent with a Kerr BH imply the presence of a LR, but they do not necessarily exclude the possibility that the merger remnant may not be a BH~\cite{Cardoso:2017njb}. 

Could the LIGO events be sourced by horizonless UCOs rather than BHs? In this work we show that UCO mergers are unlikely within a physically reasonable dynamical framework. We consider the possibility that horizonless UCOs form from the gravitational collapse of unknown forms of matter that can withstand collapse into a BH. Assuming cosmic censorship~\cite{Penrose:1969pc} and causality, such UCOs are smooth and topologically trivial~\cite{Geroch:1967fs}. For such UCOs we prove that LRs always come in pairs, one being a saddle point and the other a local extremum of an effective potential. The local extremum might be either stable or unstable, but Einstein's equations imply that instability is only possible if the UCO violates the null energy condition. Thus, UCOs formed through the collapse of reasonable (albeit exotic) matter \textit{must have a stable LR}.

It has been argued that spacetimes with a stable LR are nonlinearly unstable~\cite{Keir:2014oka,Cardoso:2014sna}. Unless these instabilities operate on time scales much longer than a Hubble time, our results imply that smooth, physically reasonable UCOs are generically unstable, and therefore that these objects are unfit as sensible observational alternatives to BHs.

%\newpage
%%%%%%%%%%%%%%%%%%%%%%%%%%%%%%%%%%%%%%%%%%%%%%%%%%%%%%%%%%%%%%%%%%%%%%%%%%%%%
\noindent{{\bf{\em Setup.}}}
%%%%%%%%%%%%%%%%%%%%%%%%%%%%%%%%%%%%%%%%%%%%%%%%%%%%%%%%%%%%%%%%%%%%%%%%%%%%%
%
%\clearpage
Various sorts of exotic compact objects have been discussed in the literature, some of which may become sufficiently compact to possess LRs. These include boson~\cite{Schunck:2003kk} and Proca stars~\cite{Brito:2015pxa}, gravastars~\cite{Mazur:2001fv}, superspinars~\cite{Gimon:2007ur} and wormholes~\cite{Visser:1995cc}. Most of these models, however, are incomplete, in the sense that no dynamical formation mechanism is known. Boson stars are an exception in this regard, because they have been shown to form dynamically (at least in spherical symmetry) from a process of gravitational collapse and cooling~\cite{Seidel:1993zk}. It is unclear whether collapse can produce ultracompact, rotating boson stars: in fact, recent numerical simulations suggest that it may not be possible to produce rotating boson stars from boson star mergers~\cite{Bezares:2017mzk}. Still, we take spherically symmetric simulations with gravitational cooling as a plausibility argument that some UCOs could form dynamically from classical (incomplete) gravitational collapse, starting from an approximately flat spacetime. The collapse stalls before the formation of an event horizon or high-curvature region, but the resulting compactness allows for LRs. Assuming causality, classical dynamical formation from an approximately flat spacetime implies, via a theorem of Geroch~\cite{Geroch:1967fs}, that the resulting spacetime is topologically trivial, so that the discussion does not apply (e.g.) to wormholes. 

Once equilibrium is attained, we assume that the UCO is described by a 4-dimensional, stationary and axisymmetric geometry. We use quasi-isotropic coordinates $(t,r,\theta,\varphi)$, adapted to the commuting azimuthal ($\partial/\partial \varphi$) and stationarity ($\partial/\partial t$) Killing vectors. We further assume that the metric is invariant under the simultaneous reflection $t\to-t$ and $\varphi\to-\varphi$. The metric functions are assumed to be everywhere smooth (apart from standard spherical coordinate singularities). No event horizon exists, and no reflection symmetry $\mathbb{Z}_2$ is required on the equatorial plane $\theta=\pi/2$. Gauge freedom is used to set $g_{r\theta}=0$, $g_{rr}>0$ and $g_{\theta\theta}>0$. To prevent closed time-like curves we require $g_{\varphi\varphi}>0$. Here and until otherwise specified we do not make assumptions on the field equations, so that the results apply to any metric theory of gravity in which photons follow null geodesics.

The Hamiltonian $\mathcal{H}=\frac{1}{2}g^{\mu\nu}p_\mu\,p_\nu=0$ determines the null geodesic flow, where $p_\mu$ denotes the photon's 4-momentum. The two Killing symmetries yield two conserved quantities: $p_t\equiv -E$ and $p_\varphi\equiv \Phi$, respectively (minus) the photon's energy and angular momentum at spatial infinity. The Hamiltonian can be split into a potential term, $V(r,\theta) \leqslant 0$, plus a kinetic term, $K\geqslant 0$: $2\mathcal{H}=K+V=0$, where
\beq
&&K\equiv g^{rr}{p_r}^2  + g^{\theta\theta}{p_\theta}^2  \ ,  \\
&&V\equiv  g^{tt}E^2 - 2g^{t\varphi}E\,\Phi + g^{\varphi\varphi}\Phi^2 \ .
\eeq
{A LR is a null geodesic with a tangent vector field that is always a linear combination of (only) the Killing vectors $\partial_t$ and $\partial_\varphi$. This implies that the momentum must satisfy} $p_r=p_\theta=\dot{p}_\mu=0$. These conditions can be reformulated using the effective potential $V$. Indeed, $\mathcal{H}=0$ implies that the following three conditions are equivalent:  
\be
V=0 \quad \Leftrightarrow \quad 
K=0 \quad \Leftrightarrow \quad 
p_r=p_\theta=0\,.
\ee
Hamilton's equations imply
\be
\dot{p}_\mu=-\left(\partial_\mu g^{rr}p_r^2 + \partial_\mu g^{\theta\theta}p_\theta^2 + \partial_\mu V\right)/2\,, 
\ee
where the dot denotes a derivative with respect to an affine parameter. Thus, at a LR ($p_r=p_\theta=\dot{p}_\mu=0$) we must have 
\begin{equation}
\quad V=\nabla V=0 \, . 
\label{lrc}
\end{equation}

The projection of a LR orbit on the configuration space $(r,\theta)$ will be simply a point, not necessarily on the equatorial plane.
Moreover, a LR will be stable (unstable) along a direction $x^\alpha$ if $\partial^2_\alpha\,V$ is positive (negative).

%%%%%%%%%%%%%%%%%%%%%%%%%%%%%%%%%%%%%%%%%%%%%%%%%%%%%%%%%%%%%%%%%%%%%%%%%%%%%
\noindent{{\bf{\em The potential functions $H_\pm$.}}}
%%%%%%%%%%%%%%%%%%%%%%%%%%%%%%%%%%%%%%%%%%%%%%%%%%%%%%%%%%%%%%%%%%%%%%%%%%%%%
The ``potential'' $V$ has the shortcoming of depending on the photon's parameters $(E,\Phi)$. Healthier potentials can be introduced as follows~\cite{Cunha:2016bjh,Cunha:2017eoe}. First, cast $V$ in terms of the covariant metric components:
\begin{equation}V=-\frac{1}{D}\left(E^2g_{\varphi\varphi} + 2E\Phi g_{t\varphi} + \Phi^2g_{tt}\right),\label{V-def}\end{equation}
where $D\equiv g^2_{t\varphi}-g_{tt}g_{\varphi\varphi}> 0$. Second, observe that $\Phi\neq 0$ at a LR. Indeed, if $\Phi=0$ and $E\neq 0$, then $V\neq 0$, and Eq.~\eqref{lrc} implies that a LR is not possible. Furthermore physical photons can not have $E=\Phi=0$, because they must satisfy $E>-\Phi\,g_{t\varphi}/g_{\varphi\varphi}$ (from the requirement that the photon's energy must always be positive for a local observer~\cite{Cunha:2016bjh}). 

Since $\Phi\neq 0$, we define the (inverse) impact parameter $\sigma\equiv E/\Phi$ and factorize $V$ as
$V=-{\Phi^2}g_{\varphi\varphi}(\sigma-H_+)(\sigma-H_-)/D$,
introducing the everywhere regular potential functions
\be
H_\pm(r,\theta)\equiv \frac{-g_{t\varphi}\pm\sqrt{D}}{g_{\varphi\varphi}}\,,
\ee
which are independent of the orbital parameters $E,\Phi$.
The LR condition $V=0$ requires that either $\sigma=H_+$ or $\sigma=H_-$. These conditions are mutually exclusive, since $\sigma=H_\pm$ implies that $\sigma-H_\mp =H_\pm - H_\mp=\pm 2{\sqrt{D}}/{g_{\varphi\varphi}}\neq 0$, and in fact they are not constraints on $H_\pm$. They simply determine the impact parameter at the LR. Thus, the LR conditions, $V=\nabla V=0$ translate, for the potentials $H_\pm$, into the sole requirement of a \textit{critical point}: $\nabla H_\pm=0$.

To infer the stability of a LR one considers the second derivatives of $H_\pm$. In particular, at a LR:
\be
\partial_\mu^2V=\pm\left(\frac{2\Phi^2}{\sqrt{D}}\right)\partial_\mu^2H_\pm\,.
\ee
Thus, the signs of $\partial_\mu^2V$ and $\pm\partial_\mu^2H_\pm$ coincide. A LR can be either a local extremum of $H_\pm$ or a saddle point. A saddle point has two proper directions with opposite stability properties, determined as the eigenvectors of the Hessian matrix at the LR; at a local extremum, both directions have the same stability properties. In particular, if both directions are stable the LR is stable, otherwise it is globally unstable.

%%%%%%%%%%%%%%%%%%%%%%%%%%%%%%%%%%%%%%%%%%%%%%%%%%%%%%%%%%%%%%%%%%%%%%%%%%%%%
\noindent{{\bf{\em LRs always come in pairs.}}}
%%%%%%%%%%%%%%%%%%%%%%%%%%%%%%%%%%%%%%%%%%%%%%%%%%%%%%%%%%%%%%%%%%%%%%%%%%%%%
We will now show that under the dynamical formation scenario we have described above, LRs of an UCO always come in pairs, with one being a saddle point and the other a local extremum of $H_\pm$. The proof relies on a simple topological argument.

Consider the vector fields ${\bf v}_\pm$, with components $v^i_\pm=\partial^i\!H_\pm$, where $i\in\{r,\theta\}$. Let $X$ be a compact, simply connected region of the ($r,\theta$) plane. Both $X$ and ${\bf v}_\pm$ are 2-dimensional. The fields ${\bf v}_\pm$ are \textit{maps} from $X$ to 2D-spaces $Y_\pm$, parameterised by the components of $v^i_\pm$. In particular, a point in $X$ where ${\bf v}_\pm$ vanishes -- a critical point of $H_\pm$, that describes a LR -- is mapped to the origin of $Y_\pm$.

For maps between manifolds such as the ones above, one can define a topological quantity, called the \textit{Brouwer degree} of the map (see e.g.~\cite{steffen2005topology,naber2000topology}), that is invariant under continuous deformations of the map. 
Consider two compact, connected and orientable manifolds $X,Y$ of equal dimension and a smooth map ${\bf f}: X \to Y$.
If ${\bf y}_0\in Y$ is a regular value of ${\bf f}$, then the set $f^{-1}({\bf y}_0)=\{{\bf x}_1,{\bf x}_2,\cdots\}$ has a finite number of points, with ${\bf x}_{n}\in X$, such that $f({\bf x}_{n})={\bf y}_0$, and the Jacobian $J_n=\det\left({\partial {\bf f}}/{\partial{\bf  x}_n}\right)\neq 0$.
The sign of $J_n$ embodies how the vector basis in $X$ projects into the basis in $Y$, and thus if the map is orientation-preserving or orientation-reversing. 
The Brouwer degree of the map ${\bf f}$ with respect to ${\bf y}_0\in Y$ is given by $\textrm{deg}({\bf f})=\sum_n \textrm{sign}(J_n)$. 
The central property of this quantity is that it does not depend on the actual choice of the regular value ${\bf y}_0$, but it is rather a topological property of the map itself. Moreover, it is invariant under homotopies, i.e. continuous deformations of the mapping. 

To apply this tool to our setup, we take the map ${\bf f}$ to be either of the vector fields ${\bf v}_\pm$; thus the maps have components $ f_\pm^i=v^i_\pm=\partial^i\!H_\pm$. 
We choose the origin of $Y$ as our reference point $y_0^i=\{0,0\}$. Then the degree of ${\bf f}_\pm$ becomes:
\be
\textrm{deg}({\bf f}_\pm)=\sum_n \textrm{sign}\left[\det\left({\partial_j\partial^i H_\pm}\right)\right]_{{\bf x}_n},
\ee
where $ \det\left(\partial_j\partial^i H_\pm\right)=g^{rr} g^{\theta\theta} \left[   \partial_r^2H_\pm \partial_\theta^2H_\pm-[\partial_{r\theta}^2H_\pm]^2\right]$ has sign $1$ ($-1$) for a local extremum (saddle point). Thus, we assign a \textit{topological charge} $w\equiv\textrm{sign}(J_n)$ to each point in $X$ where ${\bf v}_\pm$ vanishes, corresponding to a LR, and sum over all contributions to get the Brouwer degree of ${\bf v}_\pm$.

\begin{figure}[t]
\begin{center}
\includegraphics[width=0.485\textwidth]{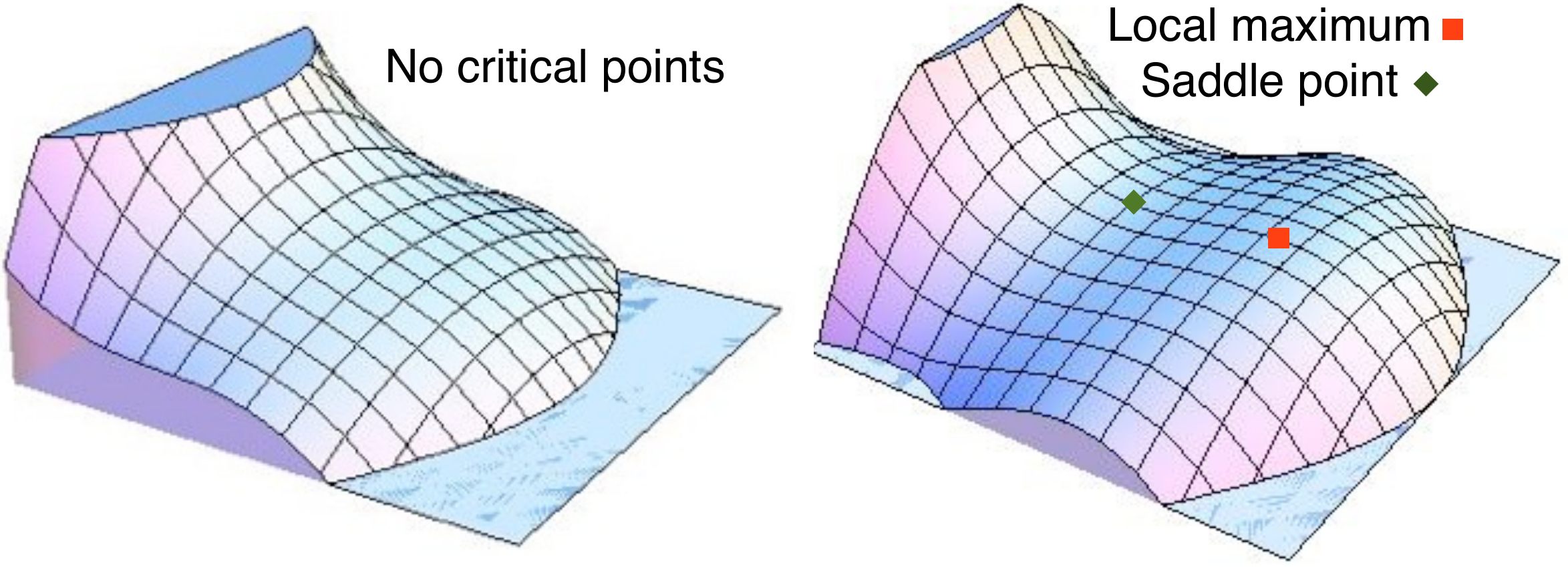}
\caption{Conservation of the Brouwer degree under a smooth deformation of a 2D map $(x,y)\rightarrow \nabla H$. We have chosen the illustrative potential $H(x,y)=x(x^2-a)-(1+x^2)y^2$, where $a$ is a \textit{local} deformation parameter that does not affect the asymptotic behavior of the map. Left panel: $a=-2$; there are no critical points and the Brouwer degree is zero. Right panel: $a=1$; there are two critical points (one local maximum and one saddle point) and the Brouwer degree is still zero.}
\label{fig_def}
\end{center}
\end{figure}

The key point is now that the degree will be preserved under a continuous deformation of ${\bf v}_\pm$, like what we have assumed will occur {as the result of} the process of (incomplete) gravitational collapse. In the initial stages of the collapse the compact object is not yet sufficiently compact to possess LRs. Thus taking X to be any $r,\theta$ domain (except for the standard spherical singularities), we have $w=0$: no points with ${\bf v}_\pm=0$ exist (cf. left panel of Fig.~\ref{fig_def} for an illustrative example). As the collapse {ends}, the ${\bf v}_\pm$ functions are smoothly deformed and LRs arise, but the total $w$ must still vanish. It follows that saddle points and local extrema of $H_\pm$ must form in pairs under a continuous deformation of the metric functions (right panel of Fig.~\ref{fig_def}). Therefore LRs must come in pairs, with one being a local extremum of $H_\pm$ and the other a saddle point. In fact, the argument applies to any spacetime that can be continuously deformed into flat spacetime.

%%%%%%%%%%%%%%%%%%%%%%%%%%%%%%%%%%%%%%%%%%%%%%%%%%%%%%%%%%%%%%%%%%%%%%%%%%%%%
\noindent{{\bf{\em Spherical Symmetry.}}}
%%%%%%%%%%%%%%%%%%%%%%%%%%%%%%%%%%%%%%%%%%%%%%%%%%%%%%%%%%%%%%%%%%%%%%%%%%%%%
So far, we have established that a smooth UCO spacetime must have at least two LRs, one of them being a local extremum of the potential, but we have not yet clarified if this extremum is a stable or an unstable LR. We will first address this question for spherically symmetric spacetimes. In this simple case we can show that such LRs are \textit{always stable}, without further assumptions. 

If the UCO spacetime is spherically symmetric, the metric can be reduced to the form:
\be
ds^2=-N(r)dt^2 + \frac{1}{g(r)}dr^2 +r^2(d\theta^2 +\sin^2\theta d\varphi^2)\ .
\ee
The functions $H_\pm$ are explicitly given in terms of the metric functions by $H_\pm=\pm{\sqrt{N}}/{(r\sin\theta)}$.
Due to symmetry, we can restrict our analysis to the equatorial plane $\theta=\pi/2$ without loss of generality; if LRs exist, they can be analyzed on this plane.

The derivatives of $H_\pm$ along $\theta$ on the equatorial plane are
\beq
&&\partial_\theta H_\pm=\mp \frac{\sqrt{N}}{r}\frac{\cos\theta}{\sin^2\theta}=0,\\
&&\partial_\theta^2H_\pm=\pm\frac{\sqrt{N}}{r}\left(\frac{1+\cos^2\theta}{\sin^3\theta}\right).
\eeq
We can then conclude that $\pm\partial_\theta^2H_\pm>0\Rightarrow \partial_\theta^2V>0.$ This implies that the effective potential is \textit{always} stable along $\theta$.

Recall that for each LR pair that is created, one LR is a local extremum of $H_\pm$, whereas the other is a saddle point. Since both LRs are stable along the $\theta$ direction, in a spherically symmetric spacetime the local extremum of $H_\pm$ must be a globally stable LR.

%%%%%%%%%%%%%%%%%%%%%%%%%%%%%%%%%%%%%%%%%%%%%%%%%%%%%%%%%%%%%%%%%%%%%%%%%%%%%
\noindent{{\bf{\em Axisymmetry.}}}
%%%%%%%%%%%%%%%%%%%%%%%%%%%%%%%%%%%%%%%%%%%%%%%%%%%%%%%%%%%%%%%%%%%%%%%%%%%%%
We now turn to the generic case of axi-symmetry (and stationarity). So far, the arguments have made quantitative use of test photon dynamics but \textit{not} of spacetime dynamics, making them independent on the equations of motion. In order to assess, in the generic axisymmetric case, if the LR that extremizes $H_\pm$ is a local maximum or minimum of $V$, we will assume Einstein's field equations (in geometrized units): $G^{\mu\nu}=8\pi\,T^{\mu\nu}$. If the energy-momentum tensor $T^{\mu\nu}$ satisfies, at every point on the spacetime, the \textit{null energy condition}
\begin{equation}
\rho\equiv T^{\mu\nu}\,p_\mu\,p_\nu\geqslant 0
\label{eq-NEC} 
\end{equation}
for any null vector $p^\mu$ (i.e. $p_\mu\,p^\mu=0$), it follows that the LR that extremizes $H_\pm$ is a local \textit{minimum} of $V$, and hence globally stable. 
To establish this result we will restrict $p_\mu$, from all the
possible null vectors, to be the 4-momentum of a null geodesic.
Moreover, we will restrict the computation of $\rho$ to the location
of a LR orbit. It will be convenient to split the spacetime
coordinates into two sets: $x^\mu=\{ x^a, x^i\}$, where
$\{x^a\}=(t,\varphi)$ and $\{x^i\}=(r,\theta)$.
We will use Greek indices for the full range of spacetime coordinates, early latin indices $(a,b,c,d)$ for the Killing coordinates $(t,\varphi)$ and middle alphabet latin indices $(i,j,k)$ for the nontrivial directions $(r,\theta)$.
With this notation, we note the following properties. In general,
$\partial_a g_{\mu\nu}=0$, $g_{ai}=0$ and $p_a={\rm constant}$.
Moreover, \textit{specifically at LRs},
$p_i=0$ and $V=p_a\,p^a=0$.

Next, we wish to compute the derivatives of $V$ and compare them with different geometrical quantities \textit{at LRs}. It will be useful to bear in mind that the metric $g_{\mu\nu}$ is block-diagonal in the $\{x^a\}$ and $\{x^i\}$ parts. Hence, e.g., $g^{ab}g_{b\mu}=\delta^a_\mu$. 
%
%\eb{$\mu$ or $i$?}
%
We start by computing the first derivatives of the potential $V$:
\begin{equation}
\partial^a V=0 \ , \qquad \partial^i V=-p^a p^b \partial^i g_{ab} \ .
\end{equation}
Looking at the Christoffel symbols $\Gamma^\mu_{ab}$ one then obtains:
\begin{equation}
%\hspace{3cm}
\frac{1}{2}\partial^\mu V= \Gamma^\mu_{ab}p^a p^b \ .\label{D1V}\end{equation}
Observe that this expression is nontrivial only for $\mu=i$.

We now need the second derivatives of $V$. A slightly lengthier computation shows that
\begin{equation}
p^a p^b\partial_i \Gamma^i_{ab}=\frac{1}{2}\partial_i\partial^i V -
2 \mathcal{B}\ ,
\label{D2V}\end{equation}
where $\mathcal{B} \equiv \left( g_{ab}p_c p_d g^{ij}\partial_i g^{ac}\partial_j g^{bd}\right)/2$.
Now we invoke Einstein's field equations to write $\rho$, defined in Eq.~(\ref{eq-NEC}), as:
\be
8\pi\rho=\left(R^{\mu\nu}-\frac{1}{2}g^{\mu\nu}R\right)p_\mu p_\nu=R_{\mu\nu}\,p^\mu\,p^\nu \ .
\ee
Equivalently, expanding the Ricci tensor, we have at a LR:
\begin{equation} 
8\pi\rho =p^a p^b \Big(\partial_i\Gamma^i_{ab}-\Gamma^\mu_{a\nu}\Gamma^\nu_{b \mu}\Big) \ . 
\label{rho2}
\end{equation}
An expression for the first term on the right hand side is provided by Eq.~\eqref{D2V}. Concerning the second term, it can be re-expressed, at a LR, as:
\begin{equation}
p^a p^b\,\Gamma^\mu_{a\nu}\Gamma^\nu_{b\mu}= -\mathcal{B} \ .
%\frac{1}{2}g_{ab}p_c p_d g^{ij}\partial_i g^{ac}\partial_j g^{bd} \ .
\label{term2}
\end{equation}
Plugging Eqs.~\eqref{term2} and~\eqref{D2V} into Eq.~\eqref{rho2} yields
\begin{equation}
8\pi\rho=\frac{1}{2}\partial_i\partial^i V - \mathcal{B} \ .
%\frac{1}{2}g_{ab}p_c p_d g^{ij}\partial_i g^{ac}\partial_j g^{bd} \ .
\label{rho3}
\end{equation}

We will now show that $\mathcal{B}=0$ at a LR. Since $p_a=$const. we can rewrite $\mathcal{B}=g_{ab}\partial^i\left(p^a\right) \partial_i\left(p^b\right)/2$, or more explicitly:
\begin{align}
2\mathcal{B}&=g^{rr}\left\{g_{tt}\left(\partial_r\dot{t}\right)^2 +2g_{t\varphi}\left(\partial_r\dot{t}\right)\left(\partial_r\dot{\varphi}\right) + g_{\varphi\varphi}\left(\partial_r\dot{\varphi}\right)^2\right\} \nonumber
\\
&+ g^{\theta\theta}\left\{g_{tt}\left(\partial_\theta\dot{t}\right)^2 +2g_{t\varphi}\left(\partial_\theta\dot{t}\right)\left(\partial_\theta\dot{\varphi}\right) + g_{\varphi\varphi}\left(\partial_\theta\dot{\varphi}\right)^2\right\} \ .
\label{b2}
\end{align}
The ``trick'' is now to write $\partial_i\dot{\varphi}$ as a function of $\partial_i\dot{t}$. Since $V=p_a p^a=-E\dot{t} +\Phi\dot{\varphi}$, we have $\partial_i V=-E\partial_i\dot{t} +\Phi\partial_i\dot{\varphi}$.
At a LR $\partial_i V=0$, and thus $\partial_i\dot{\varphi}=\left({E}/{\Phi}\right)\partial_i\dot{t}$. Returning to $\mathcal{B}$, Eq.~\eqref{b2} becomes:
\be
2\mathcal{B}=\left[g^{rr}\left(\partial_r\dot{t}\right)^2
  + g^{\theta\theta}\left(\partial_\theta\dot{t}\right)^2\right]\left[g_{tt}+2g_{t\varphi}\frac{E}{\Phi} +g_{\varphi\varphi}\frac{E^2}{\Phi^2}\right].
\ee
By comparing with Eq.~\eqref{V-def} we see that the last factor is proportional to $V$, and so it vanishes at a LR. From~\eqref{rho3} we therefore conclude that, at a LR:
\begin{equation}
\rho\equiv T^{\mu\nu}p_\mu p_\nu=\frac{1}{16\pi}\partial_i\partial^i V \ .
\label{final}
\end{equation}
This elegant and compact result informs us that the trace of the Hessian matrix of $V$ at a LR determines whether the null energy condition is violated or not. Explicitly, at a LR, $\partial_i\partial^i\,V=g^{rr}\partial^2_rV + g^{\theta\theta}\partial^2_\theta V$. Since $g^{rr}>0$ and $g^{\theta\theta}>0$, if $\partial^2_{r}V$ and $\partial^2_{\theta}V$ are both negative (positive) then the null energy condition is violated (satisfied).

{We could also consider extensions of Einstein's theory whose field equations may be written as $G^{\mu\nu}=8\pi\,T_{\rm eff}^{\mu\nu}$, where $T_{\rm eff}^{\mu\nu}$ is an \textit{effective} energy momentum tensor. Then, trivially, a similar result applies, but now the Null Energy Condition (NEC) is stated in terms of this tensor: $T_{\rm eff}^{\mu\nu}\,p_\mu\,p_\nu\geqslant 0$, with $p_\mu\,p^\mu=0$.}

%%%%%%%%%%%%%%%%%%%%%%%%%%%%%%%%%%%%%%%%%%%%%%%%%%%%%%%%%%%%%%%%%%%%%%%%%%%%%
\noindent{{\bf{\em Conclusions and remarks.}}}
%%%%%%%%%%%%%%%%%%%%%%%%%%%%%%%%%%%%%%%%%%%%%%%%%%%%%%%%%%%%%%%%%%%%%%%%%%%%%
%
It has long been suggested that ``BH mimickers'' -- horizonless ultra-compact objects of a mysterious nature and composition -- could exist in Nature. Detailed observations of celestial BH candidates in electromagnetic or gravitational radiation are expected to provide clear smoking guns to distinguish concrete models of BH mimickers from ``ordinary'' BHs. 

GWs are one of the cleanest and most pristine observables to investigate the true nature of BH candidates, in particular in the wake of the first detections by LIGO. Recent intriguing arguments imply that UCOs could mimic ordinary BHs even in the GW channel. The potential similarity between these exotic UCOs and BHs originates from the shared feature that a LR exists, together with the realization that the most distinctive GW signature of a perturbed BH (its ringdown radiation) is initially dominated by the vibrations of this LR.  

No observational evidence exists, as yet, for UCOs; but scientific open mindedness requires considering all theoretical possibilities which are not observationally excluded. If one is willing to seriously contemplate the existence of such horizonless UCOs as BH mimickers, however, one should consider them in all of their physical aspects, starting with plausible formation scenarios. Here we conservatively assumed that UCOs form from the classical (albeit incomplete) gravitational collapse of some yet unknown form of matter. This fairly unspecific assumption, together with the assumptions that the UCO is smooth and causal, led us to a compelling conclusion: if the UCO has the necessary LR to mimic a BH's ringdown, it must also have at least another LR. If the UCO is spherically symmetric, the second LR is necessarily stable, without any further requirements. In the more general (and realistic) case where the UCO is axisymmetric, the LR is stable unless the matter collapsing to form the UCO violates the null energy condition. {These results apply to a UCO spacetime that is continuously deformable into Minkowski spacetime. The impact of nontrivial topology is briefly discussed in the Supplemental Material~\footnote{See Supplemental Material for the impact of nontrivial topology, which includes Refs.~\cite{Ellis:1973yv,James:2015ima}.}.}

These generic conclusions are in agreement with UCOs studied in the literature. For instance, explicit examples where boson stars become UCOs have been considered in~\cite{Cunha:2015yba,Cunha:2016bjh,Cunha:2017eoe}: in all these cases the matter obeys the null energy condition, and indeed LRs always emerge in pairs, with one of them being stable. There is a degenerate case in which the two LRs coincide~\cite{HOD20181}.

Note that the null energy condition is relevant in a central result of general relativity, namely Penrose's singularity theorem~\cite{Penrose:1964wq}. This theorem strengthens our results: as a byproduct of Penrose's singularity theorem, there is no need to assume that our ultracompact object is horizonless. If a trapped surface were to form, the singularity theorem would imply the formation of a curvature singularity in the future evolution of the spacetime. Thus, together with the null energy condition, our assumption of smoothness \textit{implies} that the UCOs we consider are horizonless.

The existence of a stable LR allows electromagnetic or gravitational radiation to pile up in its neighbourhood. This radiation may not decay fast enough, potentially triggering a nonlinear spacetime instability~\cite{Keir:2014oka,Cardoso:2014sna}. If such instabilities are generic, UCO candidates formed from classical gravitational collapse must have astrophysically long instability time scales in order to be considered as serious alternatives to the BH paradigm. The calculation of instability time scales in nonlinear evolutions of UCOs will require numerical work that is beyond the scope of this paper.

\bigskip

% %%%%%%%%%%%%%%%%%%%%%%%%%%%%%%%%%%%%%%%%%%%%%%%%%%%%%%%%%%%%%%%%%%%%%%%%%%%%%%
\noindent{\bf{\em Acknowledgments.}}
% %%%%%%%%%%%%%%%%%%%%%%%%%%%%%%%%%%%%%%%%%%%%%%%%%%%%%%%%%%%%%%%%%%%%%%%%%%%%%%
% \begin{acknowledgments}
%
We thank Kostas Glampedakis and the anonymous referees for useful comments.
P.C. and C.H. would like to thank the University of Mississippi and the Perimeter Institute for Theoretical Physics for hospitality.  
P.C. is supported by Grant No. PD/BD/114071/2015 under the FCT-IDPASC Portugal Ph.D. program.
E.B. is supported by NSF Grants No.~PHY-1607130 and AST-1716715,
and by FCT contract IF/00797/2014/CP1214/CT0012 under the IF2014
Programme. 
E.B.'s work was performed in part at Aspen Center for Physics, which is supported by NSF Grant PHY-1607611.
C.H. acknowledges funding from the FCT-IF programme. 
This work was partially supported by the H2020-MSCA-RISE-2015 Grant No. StronGrHEP-690904, the H2020-MSCA-RISE-2017 Grant No. FunFiCO-777740 and by the CIDMA project UID/MAT/04106/2013
The authors would like to acknowledge networking support by the COST Action CA16104.

%
% \end{acknowledgments}
%%%%%%%%%%%%%%%%%%%%%%%%%%%%%%%%%%%%%%%%%%%%%%%%%%%%%%%%%%%%%%%%%%%%%%%%%%%%%%
%\vskip 5mm

\bibliographystyle{apsrev4}
\bibliography{Ref}

\end{document}